\documentclass[12pt]{amsart}

\usepackage{latexsym, color, natbib}
\usepackage{booktabs}

\usepackage{graphicx, float}
\usepackage{amsfonts, bm}

\usepackage{algorithm}
\usepackage{algpseudocode}
\usepackage{bbm}

\usepackage[colorlinks=TRUE, citecolor=blue]{hyperref}
\usepackage{amsmath, fullpage, booktabs, amsthm}
\usepackage{amsfonts, bm}

\usepackage[margin=1in]{geometry}
\usepackage{setspace} 
\newcommand{\PP}{{P}}
\newcommand{\EE}{{\mathbf{E}}}

\newcommand{\X}{{\mathbf{X}}}
\newcommand{\Z}{{\mathbf{Z}}}
\newcommand{\x}{{\mathbf{x}}}
\newcommand{\z}{{\mathbf{z}}}
\newcommand{\St}{{\mathbf{S}}}
\newcommand{\dif}{\mathrm{d}}
\newcommand{\boldtheta}{\boldsymbol{\theta}}

\begin{document}

\title{Moving-Resting Process with Measurement Error in Animal Movement Modeling}

\author{Chaoran Hu$^1$$^*$, L. Mark Elbroch$^2$, Thomas Meyer$^3$, Vladimir Pozdnyakov$^1$, and Jun Yan$^1$}

\thanks{\\
1. Department of Statistics, University of Connecticut, 215 Glenbrook Rd., U-4120, Storrs, CT 06269, U.S.A.\\
2. Panthera, 8 West 40th Street, 18th Floor, NY 10018, U.S.A.\\
3. Department of Natural Resources \& the Environment, University of Connecticut, 1376 Storrs Road, Unit 4087 Storrs, CT 06269, U.S.A.\\
* E-mail: {\tt chaoran.hu@uconn.edu}}

\label{firstpage}

\begin{abstract}
Statistical modeling of animal movement is of critical importance.
The continuous trajectory
of an animal's movements is only observed at discrete, often
irregularly spaced time points. Most existing models cannot handle the
unequal sampling interval naturally and/or do not allow inactivity
periods such as resting or sleeping. The recently proposed
moving-resting (MR) model is a Brownian motion governed by a telegraph
process, which allows periods of inactivity in one state of the
telegraph process. The MR model shows promise in modeling the movements of
predators with long inactive periods such as many felids, but the
lack of accommodation of measurement errors seriously
prohibits its application in practice. Here we incorporate measurement
errors in the MR model and derive basic properties of the
model. Inferences are based on a composite likelihood using the Markov
property of the chain composed by every other observed increments.
The performance of the method
is validated in finite sample simulation studies. Application to the
movement data of a mountain lion in Wyoming illustrates the utility of
the method.


\end{abstract}

\maketitle


\section{Introduction}
\label{sec:intro}

Statistical modeling of animal movement is of great importance in
addressing fundamental questions about space use, movement, resource
selection, and behavior in animal ecology
\citep{Hoot:John:McCl:Mora:anim:2017}.
The explosion of telemetric data on animal movement from the recent
advancements in tracking and observation technologies presents countless
opportunities and challenges \citep{Cagn:etal:anim:2010,Patt:etal:2017}.
Telemetry devices, like Global Positioning System (GPS) receivers,
can only determine an animal's position at discrete
moments in time so a continuous trajectory is never available.
Programming the device to record at a very fast fix rate could
approximate a continuous trajectory, but this is seldom done due to
battery life limitations: it is very expensive to capture and collar
an animal, so a long episodic time record is usually preferable to
a highly detailed record, especially for animals that spend a great
deal of time not moving around.
GPS receivers often produce fixes at irregularly spaced time points,
even if researchers program for regular intervals, due to environmental
factors such as satellite communication issues or sky occlusion.
As a result, discrete-time models such as the state space
model \citep[e.g.,][]{Jons:etal:2005, Patt:etal:2008, McClintock:2012}
are not realistic.
Continuous-time models based on stochastic differential equations (SDE)
\citep[e.g.,][]{Prei:etal:mode:2004, Horn:etal:2007, Bril:mode:2010} can
handle the irregular spacing naturally. Nonetheless, most existing
work assumes perpetual motion and cannot accommodate periods of inactivity.
On the time scale of most telemetry data, most animals alternate
between periods of movement (foraging) and periods of inactivity
\citep[e.g., prey handling or rest][]{Mash:etal:evid:2010,
  Ueno:etal:dopa:2012, jona:2007}. Realistic continuous-time models that
accommodate inactive periods are needed.

The recently proposed moving-resting (MR) process by \citet{Yan:etal:2014}
is a promising model to accommodate inactive periods. The MR
process is a Brownian motion governed by a telegraph or on-off process
\citep[e.g.,][]{Zack:gene:2004}. Specifically, it allows an animal to
alternate between a moving state, during which it
moves in a Brownian motion (BM), and a resting state, during which it
remains motionless. The switch between the two states is
governed by a telegraph process, where the holding time (or duration)
of each state is assumed to follow an exponential distribution. The
memoryless holding time makes the underlying state process a
continuous time Markov Chain. As a consequence, the MR process can be
analyzed with the help of hidden Markov model (HMM) tools
\citep{Capp:etal:Infe:2005}. The MR process is a first step towards more
realistic animal movement modeling with discretely observed
telemetry data where the trajectories contain evident motionless
segments. Implementation of likelihood based inferences for the MR
process based on dynamic programming \citep{Pozd:etal:2017} is
publicly available in an R package \texttt{smam} \citep{Rpkg:smam}.

The MR model does not accommodate the measurement error or noise of telemetric
devices, which is a major limitation in applying it to animal movement data.
Adding measurement error to a Brownian motion model is not
crucial as long as the noise is small in comparison to the
total standard deviation of the increments of the Brownian motion
between two consecutive time points \citep{Pozd:etal:2014}. In such
cases, discarding the noise would not produce
significant bias. The impact of the noise on inferences
about MR processes, however, is much greater.
For a given sequence of hidden states, the likelihood
is a product of both densities {\it and} probabilities. With perfect
instrumentation, if a sequence of observed
locations are exactly the same, that is, there is no change in either
the easting nor northing coordinates for a period of time resulting in
a ``flat'' piece of trajectory, then the animal is known to be motionless over the time
period spanned by the sequence. The likelihood contribution is the probability of staying in
the motionless state instead of the density of the increment at
zero. Adding even a tiny bit of noise would remove those flat pieces and,
hence, cause drastic bias in the likelihood
estimator of the parameters. One possible remedy is to round the
observed coordinates, which enforces flat pieces. The number of such
pieces, however, depends greatly on the rounding level, and there
are no obvious rules to aid researchers in choosing best levels.
A detailed illustration of the issue is given in Section~\ref{sec:mr}.

Dealing with added noise in an MR process is challenging because it
invalidates the Markov property of the joint location-state process.
The transition density from one time point to the next can in
principle be obtained from convolving the results for the MR process
\citep{Yan:etal:2014} with normally distributed noises,
although computationally very intensive. A lack of Markov property of
the joint location-state process means that the likelihood cannot be
easily formed by multiplying these transition densities. Because the
measurement errors are continuous, the dynamic programming tools of HMM
based on a finite number of hidden states \citep{Capp:etal:Infe:2005}
are not applicable. The generic simulation based inferences such as
iterated filtering \citep{ionides2011iterated, ionides2015inference}
or particle Markov chain Monte Carlo \citep{andrieu2010particle},
available in R package \texttt{pomp} \citep{king2016statistical}, are
not applicable to our investigation due to the complexity of the MR
process with measurement error.

Our contribution in this paper is a toolbox for applying the MR process with
measurement error to animal movement modeling. First, we show that
discarding the measurement error, even tiny ones, causes severe
bias in estimation, and that rounding does not provide any
satisfactory solution. To make inferences for MR process with
measurement error, we establish that, after thinning every other
observation, the remaining observations are location-state
Markov. This facilitates a composite likelihood which contains two
true likelihood components, one based on odd-numbered observations and
the other based on even-numbered observations. The true likelihood of
each component is computed with dynamic programming. The variance of
the maximum composite likelihood estimator can be estimated
through parametric bootstrap. The
validity of the approach is confirmed in a simulation study. We then
apply the approach to model the movement data of a mountain lion in
Wyoming, whose trajectory is known to have long inactive periods.
Our methods are publicly available in an R package \texttt{smam}
\citep{Rpkg:smam} with efficient C++ code.

\section{Moving-Resting Process}
\label{sec:mr}

The MR process is a Brownian motion with an infinitesimal
variance that is governed by an alternating renewal process with two
different holding times.
Let random variables $\{M_i\}_{i \geq 1}$ be independent exponential
variables with rate $\lambda_1$, and $\{R_i\}_{i \geq 1}$ be
independent exponential variables with rate $\lambda_0$.
These are the holding times. There are two possible alternating
sequences of the holding times, $(M_1, R_1, M_2, R_2, \dots)$
or $(R_1, M_1, R_2, M_2, \dots)$. Which one represents a particular
realization depends on an initial distribution. A continuous time
state process, $S(t)$, $t \geq 0$, takes only two values, 0 and 1, and
it is defined by the holding times. In particular, for
sequence $(M_1, R_1, M_2, R_2, \dots)$, if there exists $k\geq 0$ such
that
\[
  \sum_{j=1}^k(M_j+R_j)<t \mbox{ but } \sum_{j=1}^k(M_j+R_j)+M_k\geq
  t,
\]
then $S(t)=1$; otherwise, $S(t)=0$.
For sequence $(R_1, M_1, R_2, M_2, \dots)$, if there exists $k\geq 0$
such that
\[
  \sum_{j=1}^k(R_j+M_j)<t \mbox{ but }
  \sum_{j=1}^k(R_j+M_j)+R_k\geq t,
\]
then $S(t)=0$, otherwise, $S(t)=1$.
It is well-known that the state process is stationary, if the initial
probability of $\{S(0) = 1\}$ is set as
\[
  p_1 = \frac{\lambda_0}{\lambda_0 + \lambda_1},
\]
and the initial probability of $\{S(0) = 0\}$ is
set as $p_0 = 1-p_1$.

An MR process $X(t)$, $t \geq 0$, is defined by a
stochastic differential equation
\begin{equation}
  \label{eq:mr}
  \dif X(t) =
  \begin{cases}
    \sigma \dif B(t)       & \quad \text{if } S(t)=1,\\
    0  & \quad \text{if } S(t)=0,
  \end{cases}
\end{equation}
where $B(t)$ is the standard Brownian motion, and $\sigma$ is a
volatility parameter. It is important to note that
$\{X(t)\}_{t \geq 0}$ itself is not Markov,
but the location-state process $\{X(t), S(t)\}$ is a Markov process with
stationary increments.

Properties and inferences of the MR process have been
studied in \citet{Yan:etal:2014} and \citet{Pozd:etal:2017}. A key
element is the distribution of occupation times, that is, the total
time spent in the moving state by time $t$
\[
  M(t)=\int_0^tS(s)\dif s,
\]
and the total time spent in the resting state $R(t)=t-M(t)$.
Let $\PP_i(\cdot)$ be the conditional probability $\Pr(\cdot | S(0) = i)$.
\citet{Zack:gene:2004} derived computationally efficient formulas for the
following (defective) densities for $0 < w < t$:
\begin{align*}
	p_{11}(w, t) \dif w &= \PP_1(M(t) \in \dif w, S(t) = 1), \\
	p_{10}(w, t) \dif w &= \PP_1(M(t) \in \dif w, S(t) = 0), \\
	p_{01}(w, t) \dif w &= \PP_0(R(t) \in \dif w, S(t) = 1), \\
	p_{00}(w, t) \dif w &= \PP_0(R(t) \in \dif w, S(t) = 0).
\end{align*}

Having this at hand, one can derive the marginal distribution of the
increment $X(t) - X(0)$. Without loss of generality, let $X(0)$ to be
$0$, and $X(t)$ becomes the increment from time $0$ to time $t$.
Then, the joint distribution of the increment $X(t)$ and $S(t)$,
$t > 0$, is
\begin{align*}
	\PP_1(X(t) \in \dif x, S(t) = 1) &= h_{11}(x, t) \dif x, \\
	\PP_1(X(t) \in \dif x, S(t) = 0) &= h_{10}(x, t) \dif x, \\
	\PP_0(X(t) \in \dif x, S(t) = 0) &= h_{00}(x, t) \dif x + e^{-\lambda_0 t} \delta_0(x), \\
	\PP_0(X(t) \in \dif x, S(t) = 1) &= h_{01}(x, t) \dif x,
\end{align*}
where
$\delta_0(x)$ is the delta function with an atom at $0$,
$x \in \mathbb{R}$, and
$h_{ij}(x,t)$, $i, j\in \{0, 1\}$, are functions derived in
\citet{Yan:etal:2014}:
\begin{align*}
	h_{11}(x, t) &= e^{-\lambda_1t}\phi(x; \sigma^2t) + \int_{0}^{t} \phi(x;\sigma^2w)p_{11}(w,t)\dif w,\\
	h_{10}(x, t) &= \int_{0}^{t} \phi(x;\sigma^2w)p_{10}(w,t)\dif w,\\
	h_{00}(x, t) &= \int_{0}^{t} \phi(x;\sigma^2(t-w))p_{00}(w,t)\dif w,\\
	h_{01}(x, t) &= \int_{0}^{t} \phi(x;\sigma^2(t-w))p_{01}(w,t)\dif w.
\end{align*}
with $\phi(\cdot; \sigma^2)$ being the density function of normal
distribution $\mathrm{N}(0, \sigma^2)$. The marginal distribution of increment
$X(t)$ can be obtained by summing out $S(t)$ and $S(0)$, which forms
the basis of the composite likelihood in \citet{Yan:etal:2014}. The
full maximum likelihood estimation based on dynamic
programming was developed in \citet{Pozd:etal:2017}.

\begin{figure}[tbp]
  \begin{centering}
  \includegraphics[width=0.49\textwidth]{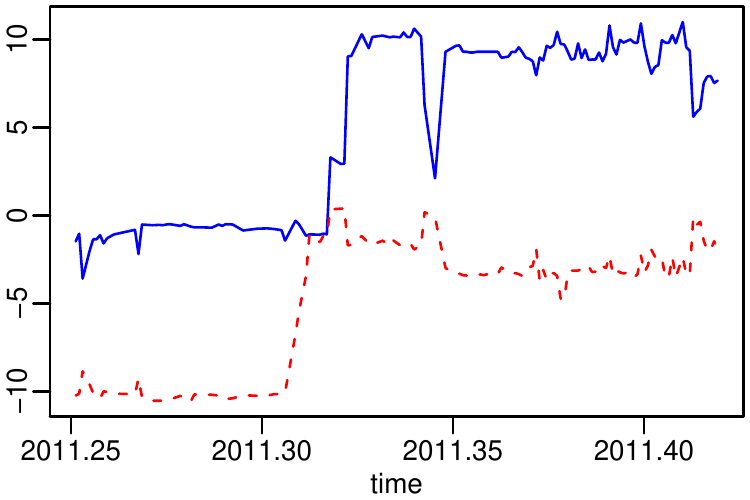}
  \includegraphics[width=0.49\textwidth]{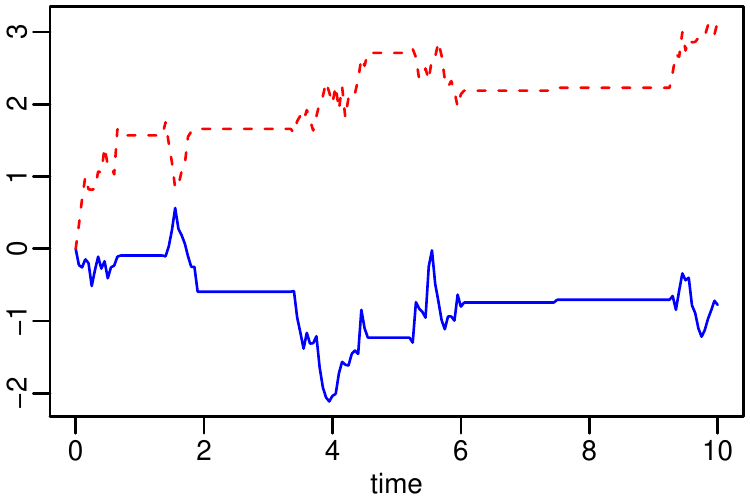}
  \includegraphics[width=0.49\textwidth]{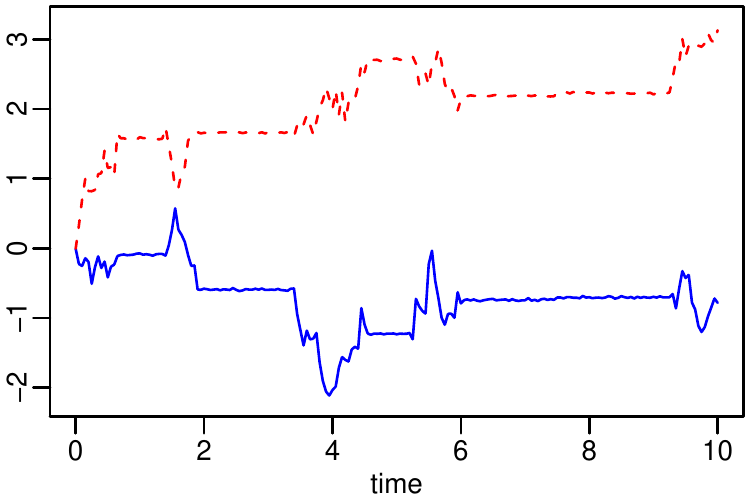}
  \includegraphics[width=0.49\textwidth]{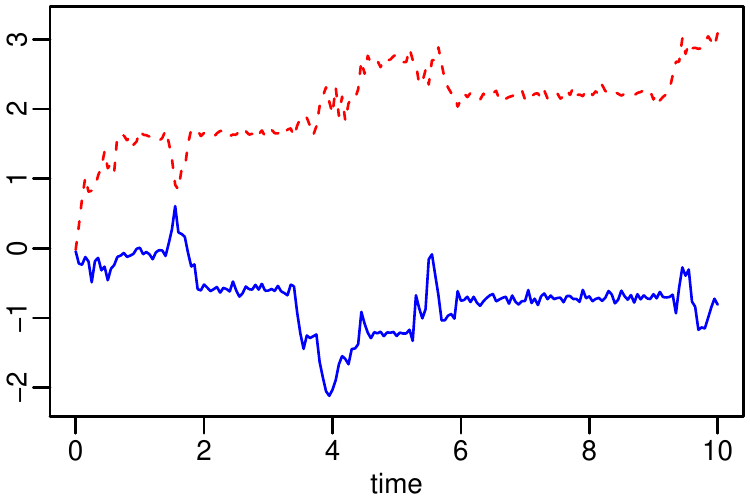}
  \caption{
    \emph{Upper left}: Actual coordinates of a female mountain lion in a
    two-month period in 2012 in the Gros Ventre mountain range, Wyoming,
    with most observations separated by 8 hours.
    The x-axis is time in years. The y-axis is departure from the starting
    point. The solid blue line is UTM easting (km) and the dashed red line
    is UTM northing (km).
    \emph{Upper right}: Coordinates of a realization from a two-dimensional
    MR process. The two coordinates are dependent because the straight line
    segments representing resting periods are shared.
    \emph{Bottom left}: Coordinates of the same realization as in upper right
    panel after adding Gaussian noise with standard deviation 0.01 km.
    \emph{Bottom right}: Coordinates of the same realization as in upper right
    panel after adding Gaussian noise with standard deviation 0.05km.
    }
  \label{fig:path}
  \end{centering}
\end{figure}

For actual animal movement data, we never observe the exact
values of $X(t)$ but only $X(t)$ with added measurement errors. For an
MR process, the probability of observing a zero increment
is strictly positive. Adding noise makes this probability zero.
Rounding can help, but
it is not trivial to come up with an appropriate rounding level.
Figure~\ref{fig:path} (upper left) shows the easting/northing coordinates
in the Universal Transverse Mercator (UTM) coordinate
system of a female mountain lion in 2012
in the Gros Ventre mountain range, Wyoming. The patterns of resting
--- places where both lines are more or less flat --- and
moving are readily apparent, which can hardly be captured by any
existing model that assumes perpetual movements.
The other three panels of Figure~\ref{fig:path} show the coordinates
of a simulated MR path without noise and with noise of two levels.
The pattern is very similar to that in the upper left panel for the female
mountain lion. The similarity is obvious, suggesting that an MR
process might be a good model, but as shown next, ignoring the noise is
disastrous in estimating the model parameters.

We demonstrate the impact of noise on estimation by a simulation
study. Consider an MR process with parameters
$\lambda_1=1\text{ hour}^{-1}$, $\lambda_0=0.5\text{ hour}^{-1}$,
$\sigma=1\text{ km}/\text{hour}^{1/2}$.
The measurement errors were independent Gaussian noise with
standard deviation $0.05\text{ km}$ ($1/20$ of $\sigma$) and $0.01\text{ km}$
($1/100$ of $\sigma$). The time intervals between consecutive
observations was 5~hours. We generated 100 datasets, each with sampling
horizon 1000 hours. The maximum likelihood estimates based on the MR process
were obtained for dataset with and without noise, where for data with
noise, three levels of rounding were considered, 10, 50, and 100
meters. Table~\ref{tab:simul:story} summarizes the parameter estimates
based on the 100 replicates. When there was no noise, the point
estimates were good, recovering the true parameters with high accuracy.
For data with noise but no rounding, the optimization did not converge
for most replications because of the Nelder-Mead simplex degeneracy
\citep{Nelder:etal:1965}. With the help of various levels of rounding,
the convergence percentage increases as the rounding becomes coarser,
and the bias decreases but remain notable. This is true even for the cases
with a noise standard deviation $0.01\text{ km}$. It is indeed unclear
how to choose an appropriate rounding level. A practical model
should handle the measurement errors directly.

\begin{table}[tbp]
  \caption{Summaries of the Influence of measurement error on MR process
    parameter estimation based on 100 replicates.
    The true parameters of the MR process were
   $\lambda_1=1\text{ hour}^{-1}$, $\lambda_0=0.5\text{ hour}^{-1}$,
   $\sigma=1\text{ km}/\text{hour}^{1/2}$. The
   measurement error was set as Gaussian noise with standard deviation
   (s.d.) $0.05$ and $0.01$. In each replicate, data were generated on
   a time horizon of $(0, 500)$ with sampling interval 1. The mean and
   standard deviation of the point estimates, along with the
   convergence percentage of the optimizations under different setups
   are reported.}
 \label{tab:simul:story}
      \centering
	\begin{tabular}{ccrrrrrrr}
          \toprule
          Gaussian noise & Rounding & Convergence & \multicolumn{2}{c}{$\hat{\lambda}_1$} &
	\multicolumn{2}{c}{$\hat{\lambda}_0$} & \multicolumn{2}{c}{$\hat{\sigma}$} \\
	\cmidrule(r){4-5}  \cmidrule(r){6-7} \cmidrule(r){8-9}
	 s.d. (km) & (km) & percentage (\%) & mean & s.d. & mean & s.d. & mean & s.d \\
	\midrule
 --- & --- & 100 & 1.00 & 0.17 & 0.50 & 0.05 & 1.00 & 0.05 \\ [2ex]
  0.05  &  --- & 7 & 348.08 & 121.53 & 2.82 & 0.08 & 5.42 & 0.98 \\
   & 0.01 & 6 & 304.83 & 127.91 & 2.80 & 0.08 & 5.05 & 1.10 \\
   & 0.05 & 14 & 240.54 & 135.93 & 2.54 & 0.11 & 4.73 & 1.34 \\
   & 0.10 & 33 & 128.79 & 91.36 & 2.06 & 0.12 & 3.94 & 1.36 \\[2ex]
  0.01 &  --- & 5 & 409.34 & 148.78 & 2.42 & 0.07 & 5.90 & 1.07 \\
   & 0.01 & 10 & 346.62 & 165.24 & 2.25 & 0.09 & 5.64 & 1.46 \\
   & 0.05 & 100 & 3.65 & 4.26 & 1.01 & 0.16 & 1.06 & 0.24 \\
   & 0.10 & 99 & 1.33 & 0.28 & 0.67 & 0.07 & 0.95 & 0.05 \\
	\bottomrule
	\end{tabular}
\end{table}

\section{Moving-Resting Process with Measurement Error}

Suppose the observations are recorded at times $t_0 = 0, t_1, \dots, t_n$.
Let  $\{\epsilon_k\}_{k=0,\dots, n}$ be independent and identically
normally distributed random variables with mean 0 and variance
$\sigma_{\epsilon}^2$. An MR process with measurement error (MRME)
$Z(t_k)$, $k = 0, 1, \dots, n$, is the superimposition of a measurement error
and the exact location:
\begin{equation}
  \label{eq:mrme}
  Z(t_k) = X(t_k) + \epsilon_k,
\end{equation}
where $X(\cdot)$ is an MR process in Equation~\eqref{eq:mr},
and $\epsilon_k$'s are independent $\mathrm{N}(0, \sigma_\epsilon^2)$ noises.

Some properties of the process
$\{Z(t_k)\}_{k=0,\dots,n}$ are in order. Obviously, it is not
Markov. Neither is the location-state process
$\{Z(t_k), S(t_k)\}_{k=0,\dots,n}$. To get a Markov process one might consider
including the measurement errors. It is true that $\{Z(t_k), S(t_k), \epsilon_k\}_{k=0,\dots,n}$
is Markov. But the cardinality of hidden states  $(S(t_k), \epsilon_k)$ is
a continuum in this case. This makes the dynamic programming approach for
computing likelihood infeasible. To address this difficulty we suggest considering the
process  $\{Z(t_{2k})-Z(t_{2k-1}), S(t_{2k})\}_{k=1,\dots,[n/2]}$. The process
is Markov, because the increment of the Brownian motion
between times $t_{2k+1}$ and $t_{2k+2}$ and measurement errors
$\epsilon_{2k+2}$ and $\epsilon_{2k+1}$  are independent of observations
collected by time $t_{2k}$. Moreover, the process has a finite set
of hidden states. This is an important property  which is used to
develop a forward algorithm for efficient computing  of a composite likelihood
in the next section. The conditional distribution of $(Z(t_{2k+2})-Z(t_{2k+1}), S(t_{2k+2}))$ (given the
observations up to time $t_{2k}$) depends only on state $S(t_{2k})$.

Here we present our derivations for a one-dimensional case. The real-world animal movement
data sets are two-dimensional. The formulas from below
can be generalized to a $d$-dimensional case. The details are given in
Appendix. All our simulations and data analysis are also performed
using two-dimensional formulas.

First, let us calculate the marginal distribution of $Z(t) - Z(0)$
(the increment of one-dimensional $\{Z(v)\}_{v \geq 0}$ from
time $0$ to time $t$).
Consider
\[
  \Delta Z(t) = Z(t) - Z(0) = X(t) - X(0) + \xi(t),
\]
where $\xi(t) \sim \mathrm{N}(0,2\sigma_{\epsilon}^2)$ is independent of process
$(X(\cdot), S(\cdot))$. Note that
\[
  Z(t_{2k+2})-Z(t_{2k+1})=X(t_{2k+2})-X(t_{2k+1})+\epsilon_{2k+2}-\epsilon_{2k+1},
\]
and
$(\epsilon_{2k+2}-\epsilon_{2k+1}) \sim \mathrm{N}(0, 2\sigma_{\epsilon}^2)$.
Without loss of generality, we assume that $X(0)=0$. Define
\[
  g_{ij}(z,t)=\PP_{i}(\Delta Z(t) \in \dif z, S(t)=j)/\dif z,
\]
where $i,j=\{1, 0\}$. Then, we get that
\begin{align*}
g_{11}(z, t) \dif z =& \PP_1(\Delta Z(t)\in \dif z, S(t)=1) \\
=& \int_{\mathbb{R}}\PP_1(X(t)+\xi(t) \in \dif z, S(t)=1, \xi(t) \in \dif x) \\
=& \int_{\mathbb{R}}\PP_1(X(t) \in \dif z - x, S(t) = 1)\phi(x; 2\sigma_{\epsilon}^2)\dif x \\
=& \int_{\mathbb{R}}h_{11}(z-x, t) \dif z \phi(x; 2\sigma_{\epsilon}^2) \dif x,
\end{align*}
where, as before, $\phi(\cdot; \sigma^2)$ is the density function of
$\mathrm{N}(0, \sigma^2)$.  Similarly, one can get that
\begin{align*}
g_{10}(z,t) =& \int_{\mathbb{R}}h_{10}(z-x, t) \phi(x; 2\sigma_{\epsilon}^2) \dif x, \\
g_{01}(z,t) =& \int_{\mathbb{R}}h_{01}(z-x, t) \phi(x; 2\sigma_{\epsilon}^2) \dif x, \\
g_{00}(z,t) =& \int_{\mathbb{R}}h_{00}(z-x, t) \phi(x; 2\sigma_{\epsilon}^2) \dif x + e^{-\lambda_0 t}\phi(z; 2\sigma_{\epsilon}^2).
\end{align*}

Next, let us denote
\[
  \tau_{ij}(t) = \PP_i(S(t)=j).
\]
It is easy to see
\begin{align*}
\tau_{01}(t) &= \PP_0(S(t)=1) \\
&= \sum_{n=0}^\infty \left[\PP\left(\sum_{k=1}^{n+1}R_k+\sum_{k=1}^nM_k \leq t\right)-\PP\left(\sum_{k=1}^{n+1}R_k+\sum_{k=1}^{n+1} M_k \leq t\right)\right] \\
&= \sum_{n=0}^\infty H(t; n, \lambda_1, n+1, \lambda_0),
\end{align*}
where $\{M_i\}_{i \geq 1}$ and
$\{R_i\}_{i \geq 1}$ are defined in~Section~\ref{sec:mr},
and a summation over an empty set is 0
and $H(t; a_1, b_1, a_2, b_2)$ is a special function involving convolutions of
independent gamma variables studied by \citet{Hu:etal:2019}.
Specifically, with $F(t; a_1, b_1, a_2, b_2)$ being the distribution
function of the sum of two independent gamma variables with parameters
$(a_1, b_1)$ and $(a_2, b_2)$, respectively,
$H(t; a_1, b_1, a_2, b_2) =
F(t; a_1, b_1, a_2, b_2) - F(t; a_1 + 1, b_1, a_2, b_2)$ can be
computed efficiently \citep[Lemma~1]{Hu:etal:2019}.
Using similar techniques, we obtain that
\begin{align*}
\tau_{10}(t) &= \sum_{n=0}^\infty H(t; n, \lambda_0, n+1, \lambda_1), \\
\tau_{00}(t) &= \sum_{n=0}^\infty H(t; n, \lambda_0, n, \lambda_1), \\
\tau_{11}(t) &= \sum_{n=0}^\infty H(t; n, \lambda_1, n, \lambda_0).
\end{align*}

Finally, we are ready to present the transition density at
$(Z(t) - Z(u), S(t))$ given $S(0)$, where $0 < u < t$,
$Z(u) = X(u) +\xi(u)$, and $\xi(u) \sim \mathrm{N}(0, \sigma_{\epsilon}^2)$ is
independent of $\xi(t)$ and process $(X(\cdot), S(\cdot))$.
Using the Markov property of the location-state process $(X(t)),S(t))$ and
the independence of the added noise, one can get that
\begin{align*}
  f(Z(t) - Z(u), S(t) = j \mid S(0) = i) =
  \sum_{k=0}^1 \tau_{ik}(u)g_{kj}(Z(t)-Z(u), t-u), \qquad 0 < u < t,
\end{align*}
where $i,j \in \{0,1\}$.

\section{Composite Likelihood Estimation}
\label{sec:CLE}

Since the full likelihood is unavailable, we resort to composite
likelihood to estimate the parameters \citep{Lind:comp:1988}.
A composite likelihood is a weighted product of likelihood segments
\[
  \mathrm{CL} = \prod_{k=1}^{K}L_{k}^{w_k},
\]
where $L_k$ is the true likelihood of the $k$-th data segment
with a non-negative weight $w_k$, $k = 1, \ldots, K$, and $K$ is the
number of segments depending on the construction of the CL. The
weights are useful, for example, in pairwise likelihood when some
pairs with stronger dependence contribute more than other pairs.
Suppose that the location-state observations are denoted as
\begin{align*}
	\Z &= (Z(t_0), Z(t_1), \dots, Z(t_n)) \\
	\St &= (S(t_0), S(t_1), \dots, S(t_n)).
\end{align*}
The observed data only contain $\Z$.
We propose two ways to construct the composite likelihood.

\subsection{Two-piece Composite Likelihood}

The likelihood of increment-state observations at even numbered time points
\begin{align*}
	\Z_{even}&= \left( Z_2, \dots, Z_{2[n/2]} \right), \\
	\St_{even}&= \left( S(t_0), S(t_2), \dots, S(t_{2[n/2]}) \right),
\end{align*}
where $Z_k = Z(t_k) - Z(t_{k-1})$ is
\begin{align*}
  &L\left(\Z_{even}, \St_{even} ; \boldtheta \right) = \nu(S(t_0)) \prod_{k=1}^{\lfloor n/2 \rfloor} f \left( Z_{2k}, S(t_{2k}) | S(t_{2k-2}) \right),
\end{align*}
in which $\lfloor a \rfloor$ is the largest integer not greater than $a$,
$\boldtheta = (\lambda_1, \lambda_0, \sigma, \sigma_\epsilon)$
and $\nu(S(t_0))$ is the initial
distribution that is assumed to be stationary.
Since $\St_{even}$ is not observed, we need to marginalize over all
possible state trajectories:
\begin{align*}
	L(\Z_{even}; \boldtheta) = \sum_{S(t_0), S(t_2), \dots, S(t_{2[n/2]}) \in \{0, 1\}} L\left(\Z_{even},\St_{even} ; \boldtheta \right).
\end{align*}
The cardinality of the set of the state trajectories is $2^{[n/2]+1}$,
which makes the direct summation infeasible for even moderate $n$.
It can, however, be tackled with the help of dynamic programming,
specifically, by the forward algorithm.

First, let us define the forward variables by
\begin{align*}
  \alpha(\Z_{even}(t_{2k}), S(t_{2k}), \boldtheta) =
  &\sum_{S(t_0), S(t_2), \dots, S(t_{2k-2}) \in \{0, 1\}} \nu(S(t_0)) \\
  &\times \prod_{j=1}^k f \left( Z(t_{2j}) - Z(t_{2j-1}), S(t_{2j}) \mid S(t_{2j-2}) \right),
\end{align*}
where $\Z_{even}(t_{2k}) = (Z(t_0), Z(t_2),\dots, Z(t_{2k}))$,
$k = 1, \dots, [n/2]$, and
the initial forward variable
$\alpha \left( \Z_{even}(t_0), S(t_0), \boldtheta \right)=\nu(S(t_0))$.
Then, it is easy to see that the forward variables satisfy the
following recursive relationship:
\begin{align*}
  \alpha \left( \Z_{even}(t_{2k+2}), S(t_{2k+2}), \boldtheta \right) =
  & \sum_{ S(t_{2k}) \in \{0,1\} } \alpha \left( \Z_{even}(t_{2k}), S(t_{2k}), \boldtheta \right)\\
  &\times f \left( \Z_{even}(t_{2k+2}) - \Z_{even}(t_{2k+1}), S(t_{2k+2}) | S(t_{2k}) \right).
\end{align*}
This allows us to compute the likelihood in linear time with respect
to $n$, because
\[
  L(\Z_{even}; \boldtheta)=\sum_{ S(t_{2[n/2]}) \in \{0,1\} } \alpha
  \left( \Z_{even}(t_{2[n/2]}), S(t_{2[n/2]}), \boldtheta \right).
\]

Now, when the sample size $n$ is large, the series of multiplications
may cause underflow problems where some terms are too small to be
distinguished from zero by a computer. A normalized forward algorithm
addresses the underflow problem.
More specifically, let us introduce the normalized forward variables as
\[
  \bar{\alpha} \left( \Z_{even}(t_{2k}), S(t_{2k}), \boldtheta \right) = \frac{\alpha \left( \Z_{even}(t_{2k}), S(t_{2k}), \boldtheta \right)}
  {L \left( \Z_{even}(t_{2k}); \boldtheta \right)},
\]
and let
\[
  \rho \left( \Z_{even}(t_{2k+2}); \boldtheta \right) = \frac{L \left(
      \Z_{even}(t_{2k+2}); \boldtheta \right)}{L \left(
      \Z_{even}(t_{2k}); \boldtheta \right)}.
\]
Then, the update formulas for normalized forward variable
$\bar{\alpha} \left( \Z_{even}(t_{2k}), S(t_{2k}), \boldtheta \right)$ and \\
$\rho \left( \Z_{even}(t_{2k+2}); \boldtheta \right)$ are
\begin{align*}
	\bar{\alpha} \left( \Z_{even}(t_{2k+2}), S(t_{2k+2}), \boldtheta \right)
               =& \frac{1}{\rho \left( \Z_{even}(t_{2k+2}); \boldtheta \right)} \\
	           &\times \sum_{ S(t_{2k}) \in \{0,1\} } \bar{\alpha} \left( \Z_{even}(t_{2k}), S(t_{2k}), \boldtheta \right) \\
               &\times f \left( \Z_{even}(t_{2k+2}) - \Z_{even}(t_{2k+1}), S(t_{2k+2}) | S(t_{2k}) \right),
\end{align*}
and
\begin{align*}
	\rho \left( \Z_{even}(t_{2k+2}); \boldtheta \right)
               =& \sum_{ S(t_{2k+2}), S(t_{2k}) \in \{0,1\} } \bar{\alpha} \left( \Z_{even}(t_{2k}), S(t_{2k}), \boldtheta \right)\\
               &\times f \left( \Z_{even}(t_{2k+2}) - \Z_{even}(t_{2k+1}), S(t_{2k+2}) | S(t_{2k}) \right).
\end{align*}
Finally, the likelihood function is given by
\begin{align}\label{eq:llk-mrme}
	\log L \left( \Z_{even}(t_{2k}); \boldtheta \right) = \sum_{k=1}^{[n/2]} \log \rho \left( \Z_{even}(t_{2k}); \boldtheta \right).
\end{align}

In a similar fashion, one can compute the likelihood
$L(\Z_{odd}; \boldtheta)$ of the observed
increments at the odd time points
$\Z_{odd}= \left( Z_1, Z_3, \dots, Z_{2[(n+1)/2]-1} \right)$.
Adding two log-likelihoods together we get the following composite
log-likelihood:
\begin{align}\label{eq:cllk}
  \mathrm{CL} ( Z(t_0), \dots, Z(t_n); \boldtheta)=
  \log  L(\Z_{even}; \boldtheta) +
  \log  L(\Z_{odd}; \boldtheta) .
\end{align}
Each piece in~\eqref{eq:cllk} is a true log likelihood
for about a half of the observations.
The maximum composite likelihood estimator (MCLE) of $\boldtheta$ is
the maximizer $\hat\boldtheta$ of~\eqref{eq:cllk}.

\subsection{Marginal Composite Likelihood}

The second approach is to use the one-step transition density with the
dependence between two consecutive increments ignored.
If $\St$ were observed, for $i, j \in \{0, 1\}$, the likelihood of
each pair of consecutive location-state observations
\[
  (\{Z(t_{k-1}), S(t_{k-1})=i\}, \{Z(t_{k}), S(t_{k})=j\})
\]
is
\[
  \nu (S(t_{k-1})=i) g_{ij}(Z(t_k)-Z(t_{k-1}), t_k-t_{k-1}),
\]
where $\nu(\cdot)$ is the stationary distribution of state process $\{S(t)\}_{t\geq 0}$.
Since $\St$ is unobserved, the likelihood of
$(Z(t_{k-1}), Z(t_k))$ is
\[
  \sum_{j=0}^1 \sum_{i=0}^1 \nu (S(t_{k-1})=i) g_{ij}(Z(t_k)-Z(t_{k-1}),
  t_k-t_{k-1}).
\]
The marginal composite log-likelihood is
\begin{align*}\label{eq:cllk_naive}
	\mathrm{CL^*}((Z(t_0), \dots, Z(t_n)); \boldtheta) = \sum_{k=1}^n \log \left[ \sum_{j=0}^1 \sum_{i=0}^1 \nu (S(t_{k-1})=i) g_{ij}(Z(t_k)-Z(t_{k-1}),
	t_k-t_{k-1}) \right].
\end{align*}
Since the dependence among the increments is discarded, the resulting
estimator is expected to be less efficient if the dependence is stronger.

\subsection{Variance Estimation}
To make inferences about $\boldtheta$, we need the variance of
$\hat\boldtheta$. It can be estimated by parametric bootstrap with the
time points fixed easily because simulating from the MRME process
is simple. The general approach of parametric bootstrap is given
as Algorithm~\ref{alg:pb}.


\begin{algorithm}
	\caption{Estimating standard error from parametric bootstrap}\label{alg:pb}
	\begin{algorithmic}[0]
		\State \textbf{input:} Observed data; number of resampling $M$.
	\State $\cdot$ Fit model to get the parameter estimates;
	\For{$m=1$ to $M$}
	\State $\cdot$ Use the estimated parameters to generate a bootstrap sample on the observed time grids;
 	\State $\cdot$ Fit model to the bootstrap sample to get bootstrap estimate;
	\EndFor
	\State $\cdot$ Make inference based on empirical distribution from these bootstrap estimators.
	\end{algorithmic}
\end{algorithm}

Alternatively, we can estimate the variance by inverting the
observed Godambe information matrix \citep{godambe:1960}
\begin{align*}
	G(\boldtheta)=H(\boldtheta)J(\boldtheta)^{-1}H(\boldtheta),
\end{align*}
where
\[
  H(\boldtheta)=\EE \left[ -\frac{\partial^2}{\partial \boldtheta^2} \mathrm{CL}
    ((Z(t_0), \dots, Z(t_n)); \boldtheta) \right],
\]
and
\[
  J(\boldtheta)=\mathrm{Var} \left[\frac{\partial}{\partial \boldtheta} \mathrm{CL}
    ((Z(t_0), \dots, Z(t_n)); \boldtheta) \right].
\]
Practically, $H(\boldtheta)$ is estimated by the Hessian matrix
of the negative composite likelihood evaluated at $\hat\boldtheta$.
Calculation of $J(\boldtheta)$ is more difficult as there is no
replicated data to estimate this variance. Parametric bootstrap can be
applied to evaluate $J(\boldtheta)$ as the empirical variance of
gradient of composite likelihood from a large number of bootstrap
samples. Finally, $\mathrm{Var}(\hat\boldtheta)$ can be obtained by
the inverse of $G(\hat\boldtheta)$ \citep[e.g.,][]{cris:2011}. This
approach, however, did not perform as well as the parametric bootstrap
approach in our numerical studies.

\section{Simulation Study}
\label{sec:sim}

We ran three simulation studies to check the performance of
the MCLE based on both the marginal
composite likelihood and the two-piece composite likelihood.
The objective of the studies were threefold:
(1) to see if the procedures successfully recovered the model
parameters, (2) to verify that standard errors could be obtained with
the help of parametric bootstrap, and (3) to
compare the performance of the marginal method to the two-piece one.

In Study~1, we generated movement data using the MRME model described by
equation~\eqref{eq:mrme}.
The model parameters were set to be $\lambda_1 = 1$,
$\lambda_0 = 0.5$, $\sigma = 1$, and
$\sigma_\epsilon \in \{0.01, 0.05\}$. This is the same setup that was used for
simulations in Section~\ref{sec:mr}. For each configuration, we
generated $200$ two-dimensional datasets on a time grid from~0
to~1000, with sampling interval~5. The resulting data has
length $n =200$.

\begin{figure}[tbp]
  \begin{centering}
  \fbox{\includegraphics[width=0.48\textwidth]{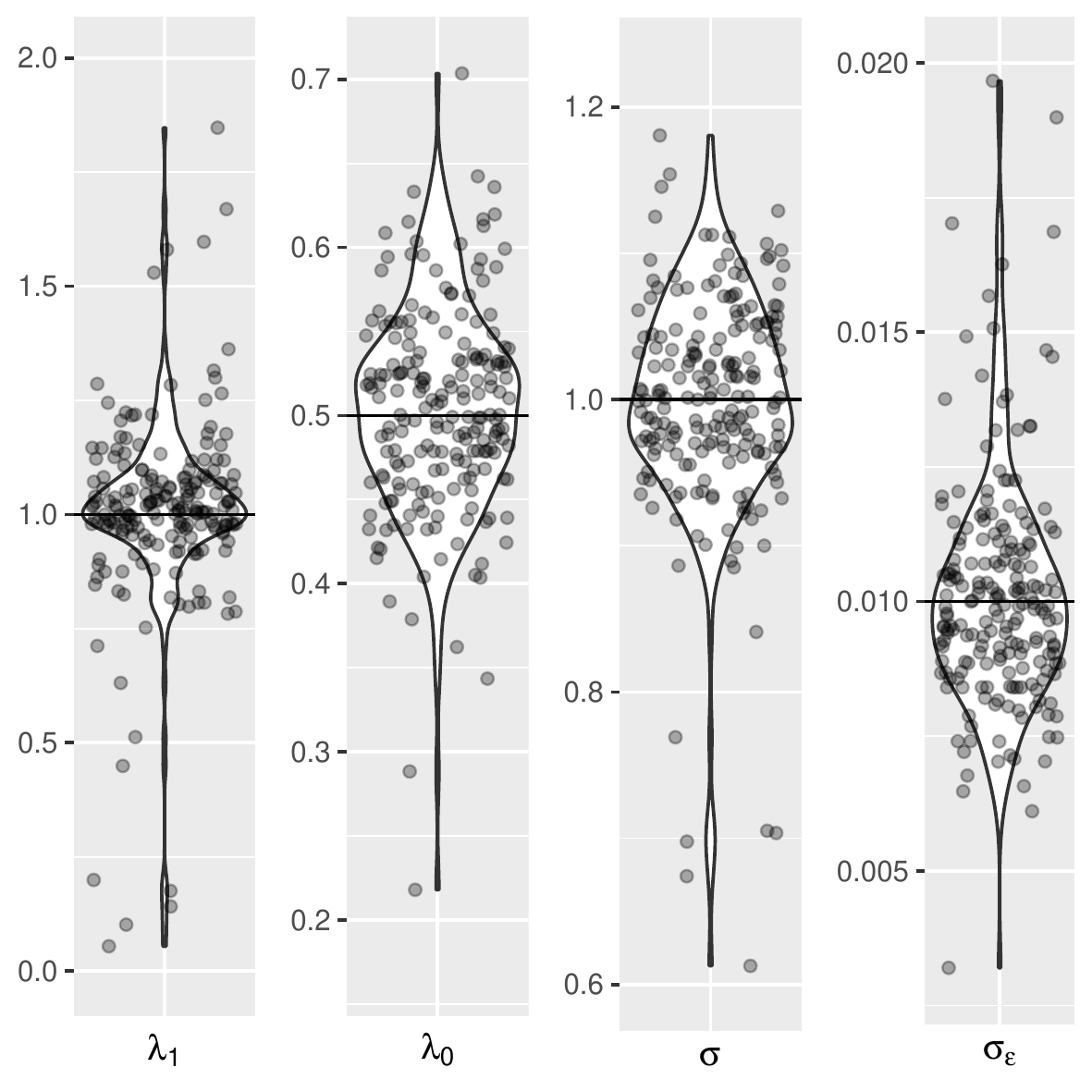}}
  \fbox{\includegraphics[width=0.48\textwidth]{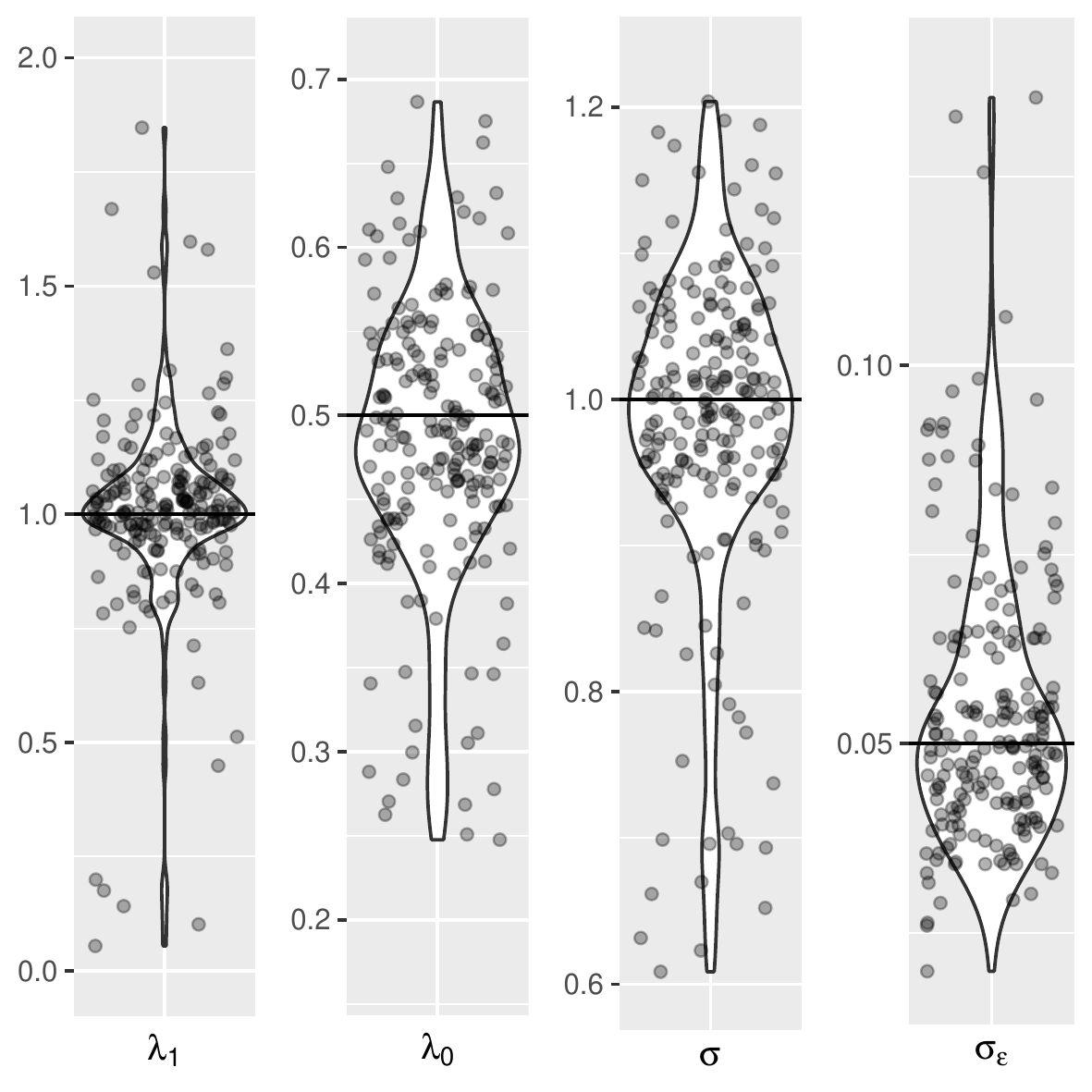}}
  \fbox{\includegraphics[width=0.48\textwidth]{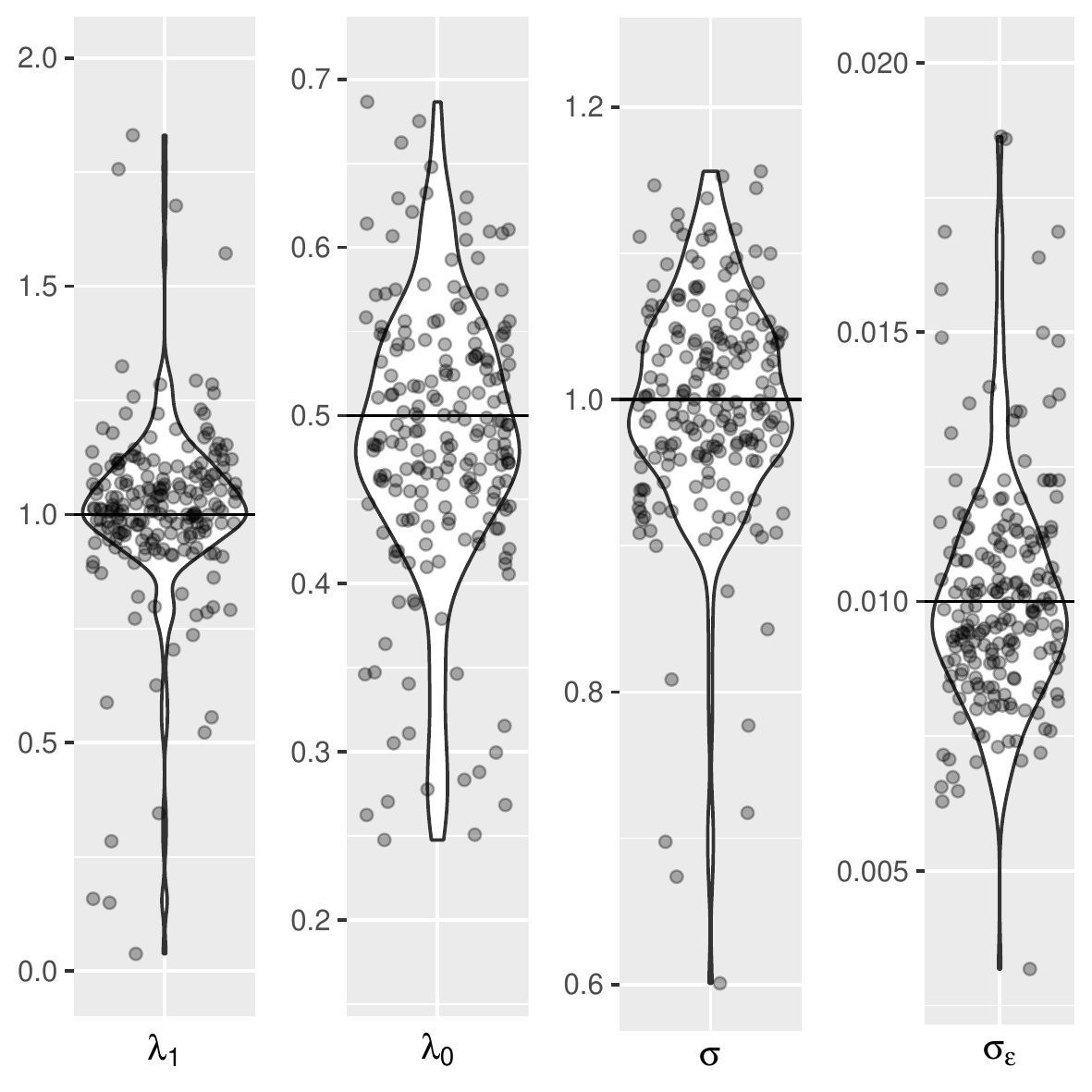}}
  \fbox{\includegraphics[width=0.48\textwidth]{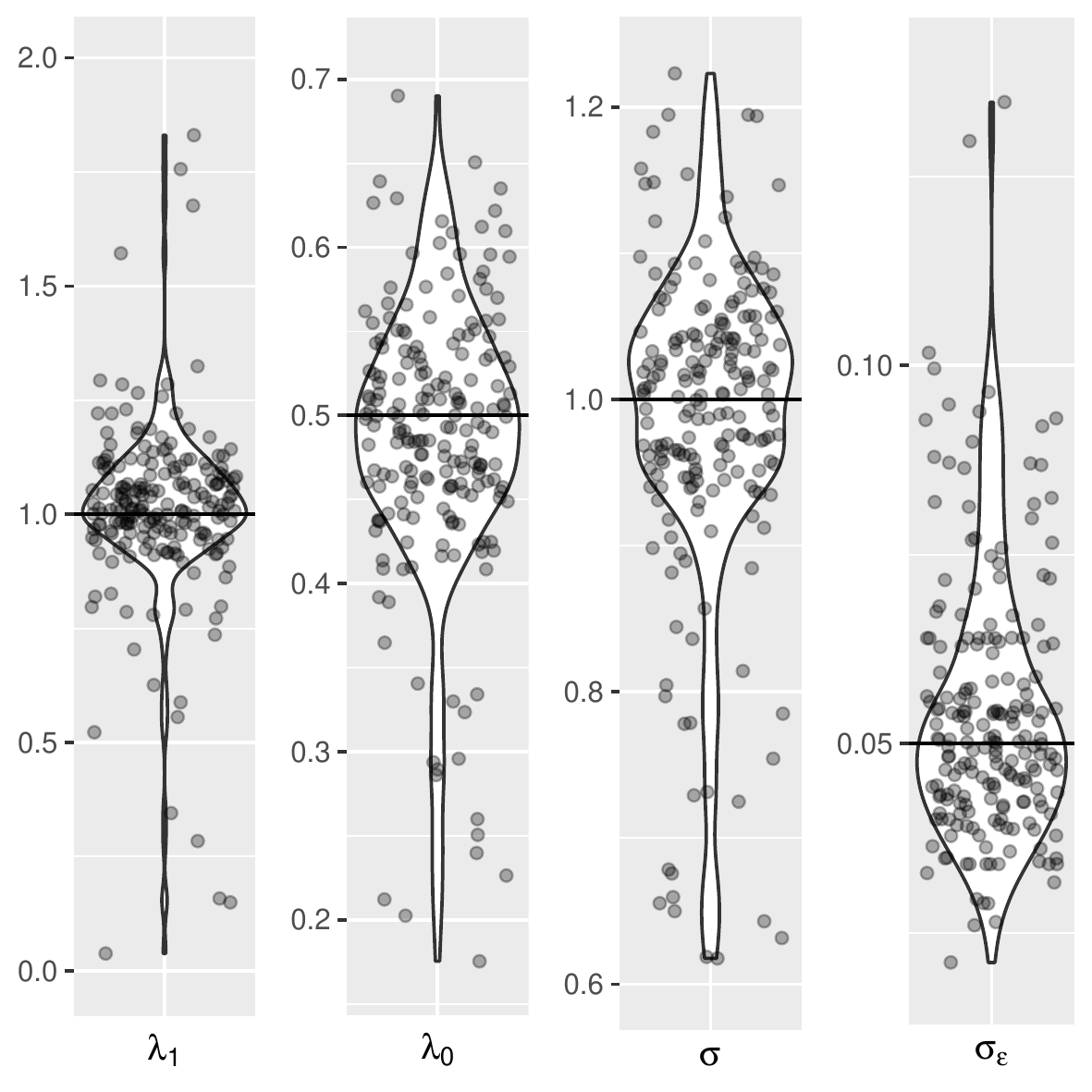}}

  \caption{
    Violin plots of the MCLE in Study~1 with two-piece method (\emph{top})
    and marginal method (\emph{bottom}) from 200 replicates. The horizontal bar in
    each panel is the true parameter value $\lambda_1=1$,
    $\lambda_0=0.5$, $\sigma=1$, and $\sigma_\epsilon=0.01$ (\emph{left})
    and $0.05$ (\emph{right}). The number of replications is 200.
    }
  \label{fig:simul}
  \end{centering}
\end{figure}

Figure~\ref{fig:simul} presents the violin plots of the MCLE in Study~1
of the 200~replicates in comparison to the true
values of the four parameters. Violin plots are similar to box plots
with a rotated kernel density plot on each side. The horizontal bars
in the panels are the true parameter values. For each parameter, the true value
lies in the bulk part of the violin plot.  This indicates that the true parameters
are recovered well by both MCLE methods.  The left and right panels represent
different
sized additional measurement errors. The variation of the estimates in the case
of $\sigma_\epsilon = 0.01$ is smaller than that in the
case of $\sigma_\epsilon = 0.05$, which is expected.

The second simulation study addresses the problem of estimating of
standard errors for both MCLE procedures via parametric bootstrap.
The sampling horizon (the length of observation
window) was set to two levels, 200 and 500 time units. The
sampling intervals (the inverse sampling frequency) also had two levels,
1 and 5 time units. The parameters of MRME process were: $\lambda_1=1$,
$\lambda_0=0.5$, $\sigma=1$, and $\sigma_\epsilon=0.01$.
Table~\ref{tab:simul:mrme4} (upper block) summarizes the results.
Once again we can see that both the marginal method and the two-piece
method recover the true parameters well. Their empirical standard errors
are similar, suggesting that the two methods have comparable efficiency
for these setups. Moreover, the standard errors were estimated
by the parametric bootstrap procedure with 50 replications.
The estimated standard errors are reasonably close to empirical ones.
The coverage rates of the 95\% bootstrap confidence intervals are as
low as 81\% for $\lambda_1$ for the case with sampling interval~5 and
sampling horizon~200. As the sampling interval decreases and the
sampling horizon increases, the coverage rates get reasonably close to
the nominal level.

Let us make a few remarks on the influence of sampling horizon
and sampling interval on the efficiency of estimation.
When the sampling interval is held fixed but the sampling horizon is 2.5 times
longer, the ESE (empirical standard error) seems to
be $\sqrt{2.5}(=1.58)$ times smaller for most
parameter estimates. The longer sampling horizon covers
more moving-resting cycles ($M_j + R_j$), and provides more information on
both mobility and measurement error parameters. If we fix the sampling
horizon and increase sampling frequency by reducing the sampling
interval, however, the ESE of only $\hat{\sigma}$ reduces in
proportion to the square root of the number of observations. Theoretically, if one can
take observations nearly continuously, the mobility parameter
$\sigma$ can be estimated with absolute accuracy. Increasing sampling frequency
also improves the estimation of $\lambda$s to a certain
degree, but improves the estimation of $\sigma_{\epsilon}$ drastically.
The results show the difference between the domain expansion
asymptotics and the in-fill asymptotics.

\begin{table}[tbp]
	\caption{Summaries of Studies~2 and~3: average estimator (EST), empirical standard
	error (ESE), average parametric bootstrap standard error (ASE),
	and coverage rate (CR) of $95\%$ large-sample confidence interval of
	MCLE with the two-piece method and the marginal method. The
        number of replications is 200.}
	\label{tab:simul:mrme4}
	\begingroup
	\setlength{\tabcolsep}{3.8pt} 
	\begin{tabular}{lllrrrrrrrrr}
		\toprule
		\multicolumn{1}{r}{Sampling} &
		\multicolumn{1}{r}{Sampling} &
		\multicolumn{1}{r}{Parameter} & \multicolumn{1}{r}{True} &
		\multicolumn{4}{c}{Two-piece method} & \multicolumn{4}{c}{Marginal method} \\
		\cmidrule(r){5-8} \cmidrule(r){9-12}
		horizon & interval & & value & EST & ESE & ASE & CR & EST & ESE & ASE & CR \\
		\midrule
		\multicolumn{12}{c}{\it Study 2}\\
200 & 5 & $\lambda_1$ & 1.0 & 0.961 & 0.546 & 0.508 & 0.80 & 0.982 & 0.570 & 0.516 & 0.81 \\
   &  & $\lambda_0$ & 0.5 & 0.493 & 0.169 & 0.240 & 0.93 & 0.485 & 0.211 & 0.251 & 0.89 \\
   &  & $\sigma$ & 1.0 & 0.966 & 0.189 & 0.163 & 0.80 & 0.970 & 0.194 & 0.166 & 0.82 \\
   &  & $\sigma_\epsilon (\times 10^{-2})$ & 1.0 & 1.145 & 0.710 & 0.651 & 0.91 & 1.136 & 0.665 & 0.648 & 0.92 \\ [2ex]
   & 1 & $\lambda_1$ & 1.0 & 1.104 & 0.394 & 0.386 & 0.94 & 1.057 & 0.410 & 0.414 & 0.88 \\
   &  & $\lambda_0$ & 0.5 & 0.502 & 0.093 & 0.092 & 0.96 & 0.488 & 0.109 & 0.109 & 0.92 \\
   &  & $\sigma$ & 1.0 & 1.011 & 0.084 & 0.088 & 0.93 & 1.001 & 0.089 & 0.097 & 0.92 \\
   &  & $\sigma_\epsilon (\times 10^{-2})$ & 1.0 & 1.002 & 0.070 & 0.068 & 0.94 & 1.002 & 0.069 & 0.068 & 0.93 \\ [2ex]
500 & 5 & $\lambda_1$ & 1.0 & 1.020 & 0.362 & 0.342 & 0.92 & 1.009 & 0.354 & 0.335 & 0.91 \\
   &  & $\lambda_0$ & 0.5 & 0.512 & 0.101 & 0.111 & 0.95 & 0.509 & 0.106 & 0.115 & 0.95 \\
   &  & $\sigma$ & 1.0 & 0.982 & 0.122 & 0.114 & 0.91 & 0.978 & 0.119 & 0.112 & 0.92 \\
   &  & $\sigma_\epsilon (\times 10^{-2})$ & 1.0 & 1.046 & 0.356 & 0.376 & 0.92 & 1.045 & 0.362 & 0.357 & 0.90 \\ [2ex]
   & 1 & $\lambda_1$ & 1.0 & 1.036 & 0.240 & 0.224 & 0.93 & 0.983 & 0.224 & 0.223 & 0.88 \\
   &  & $\lambda_0$ & 0.5 & 0.508 & 0.060 & 0.058 & 0.96 & 0.495 & 0.064 & 0.062 & 0.92 \\
   &  & $\sigma$ & 1.0 & 1.008 & 0.060 & 0.055 & 0.92 & 0.997 & 0.067 & 0.058 & 0.90 \\
   &  & $\sigma_\epsilon (\times 10^{-2})$ & 1.0 & 0.998 & 0.044 & 0.042 & 0.94 & 0.998 & 0.044 & 0.042 & 0.93 \\
       \midrule
          \multicolumn{12}{c}{\it Study 3}\\
160 & 0.8 & $\lambda_1$ & 1.0 & 1.139 & 0.440 &  &  & 1.069 & 0.549 &  &  \\
   &  & $\lambda_0$ & 0.1 & 0.106 & 0.030 &  &  & 0.096 & 0.039 &  &  \\
   &  & $\sigma$ & 1.0 & 1.013 & 0.119 &  &  & 0.991 & 0.140 &  &  \\
   &  & $\sigma_\epsilon (\times 10^{-2})$ & 1.0 & 0.998 & 0.045 &  &  & 0.998 & 0.045 &  &  \\ [2ex]

200 & 0.1 & $\lambda_1$ & 1.0 & 1.031 & 0.238 &  &  & 1.132 & 0.329 &  &  \\
   &  & $\lambda_0$ & 0.1 & 0.102 & 0.024 &  &  & 0.109 & 0.026 &  &  \\
   &  & $\sigma$ & 1.0 & 1.002 & 0.040 &  &  & 1.007 & 0.043 &  &  \\
   &  & $\sigma_\epsilon (\times 10^{-2})$ & 1.0 & 0.999 & 0.015 &  &  & 0.999 & 0.015 &  &  \\
		\bottomrule
	\end{tabular}
	\endgroup
\end{table}

Finally, let us note that the performance of the two methods is similar in both
simulation studies described above. That was surprising because the marginal
method basically ignores the dependence of the MRME process and treats
increments as if they are independent. One possible explanation is that
when the distance between two consecutive observation is relatively long, then
the dependence between two consecutive increments of the MRME process is weaker.
To illustrate this point, one can calculate the correlation
of absolute values of
consecutive increments via simulation by employing the auto-correlation function
with lag 1 (ACF(1)) and lag 2 (ACF(2)). For example, for the same
parameter set as in the above simulations and a long time horizon 100,000,
both ACF(1) and ACF(2) for the sampling interval~5 are very close to~0
from a Monte Carlo study. For the sampling interval 0.1, however,
they are 0.46 and 0.40, respectively. Our third simulation study was
based on this design. We also considered sampling interval
0.8 in the simulation because it has a relatively large ACF(1), 0.23, and
a significantly smaller ACF(2), 0.07.
The results of Study~3 with these small sampling intervals are presented in
Table~\ref{tab:simul:mrme4} (lower block). It is clear that
the two-piece procedure is preferable for datasets with shorter sampling intervals
(more frequent observations).

%

\section{Movement of a Mountain Lion}

\begin{figure}[tbp]
  \begin{centering}
  \includegraphics[width=0.49\textwidth]{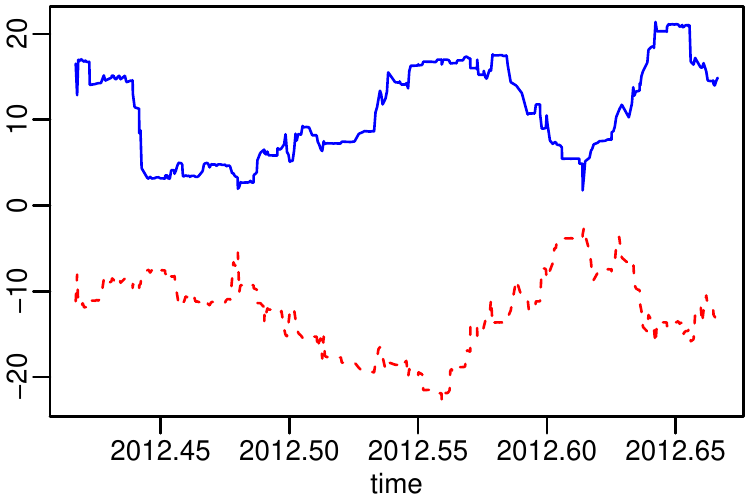}
  \includegraphics[width=0.49\textwidth]{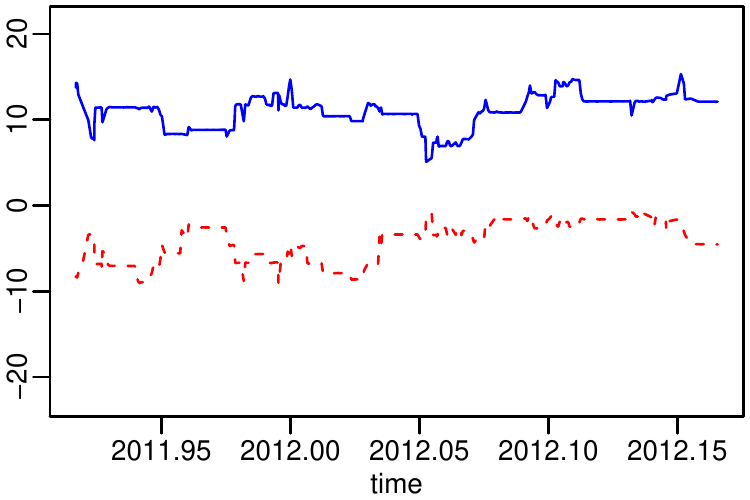}
  \caption{
    Actual coordinates of a female mountain lion in the Gros Ventre
    mountain range, Wyoming. The $x$-axis is time in years. The $y$-axis
    is departure from the starting point. The solid blue line is UTM
    easting (km) and the dashed red line is UTM northing (km).
    \emph{Left:} Summer period data, from June 1, 2012 to August 31, 2012.
    \emph{Right:} Winter period data, from December 1, 2011 to February 29, 2012.
    }
  \label{fig:catseason}
  \end{centering}
\end{figure}

The MRME model was applied to GPS data collected on a mature
female mountain lion living in the Gros Ventre
Mountain Range near Jackson, Wyoming. The data were collected
by a code-only GPS wildlife tracking collar from 2009 to 2012. The
collar was programmed to record locations every 8 hours, but the
actual sampling intervals were irregular. In the Grand Teton and
Gros Ventre mountains of Wyoming, deep winter snows ensure that
mountain-lion movements differ across seasons \citep{Elbroch:etal:2013}.
So, we fitted a MRME model to the summer data (from June 1, 2012 to
August 31, 2012) and winter data (from December 1, 2011 to February
29, 2012), respectively. These two periods of data were plotted
as Figure~\ref{fig:catseason}. The summer data had an average sampling
interval of 5.46 hours with standard deviation 5.14 hours, ranging
from 0.5 hours to 32 hours. The average sampling interval was 5.58 hours
in the winter data, with standard deviation 4.09 hours and range
from 0.5 hours to 25 hours. The summer data has 401 observations and
winter data has 392 observations.

The estimates based on two-piece composite likelihood
for summer data are $\hat{\lambda}_1 = 2.841$ $\text{hour}^{-1}$,
$\hat{\lambda}_0 = 0.179$ $\text{hour}^{-1}$, $\hat{\sigma} = 1.335$ $\text{km}/\text{hour}^{1/2}$
and $\hat{\sigma_\epsilon} = 0.019$ $\text{km}$. On average, this
lion was moving 0.352 hours for each 5.587 hours
resting, and, if she kept moving for (exactly) one full hour,
the average deviation from the initial position is $1.335$ $\text{km}$ in both
directions (northing and easting). Compared to the summer data, the
estimates for winter data are $\hat{\lambda}_1 = 6.225$ $\text{hour}^{-1}$,
$\hat{\lambda}_0 = 0.118$ $\text{hour}^{-1}$, $\hat{\sigma} = 1.506$ $\text{km}/\text{hour}^{1/2}$
and $\hat{\sigma_\epsilon} = 0.009$ $\text{km}$, so, during the winter period, she
spent $51.7\%$ more time staying in place and $54.4\%$ less time moving.
The estimate of $\sigma_\epsilon$ (standard deviation of
Gaussian noise) indicates that the GPS tracking collar had about 10-
to 20-meters of measurement error, and the error is twice as high in
summer than in winter, which is consistent with
the report by \citet{Owari:2009}. However, the variability
observed by \citet{Owari:2009} was primarily due to sky obstruction
from broadleaf tree canopy, which cannot be the case for the
conifer-dominated mountains in Wyoming. Instead, these differences
in GPS error likely reflect the fact that mountain lions in the
study area select thicker vegetation with shade and cover in which
to bed in the summer and more open terrain with southern aspects
in rugged terrain, which catch sun and provide thermoregulatory
benefits in the winter \citep{kusler:2017}. The results from the
marginal method are similar
(Table~\ref{tab:realdata}) except that the estimate of the rate
parameter $\lambda_1$ of the moving state is noticeably smaller. Based
on the comparison between the two methods in the simulation study, our
discussion used the results from the two-piece method.

\begin{table}[tbp]
	\caption{Analysis results for mountain lion movement data. Point
	estimates (EST) from both two-piece method and marginal method are
	reported. Standard error of point estimates are evaluated by
	parametric bootstrap (SE).}
	\label{tab:realdata}
	\centering
	\begingroup
	\begin{tabular}{llrrrr}
		\toprule
		\multicolumn{1}{l}{Season} & \multicolumn{1}{l}{Parameter} &
		\multicolumn{2}{c}{Two-piece method} & \multicolumn{2}{c}{Marginal method} \\
		\cmidrule(r){3-4} \cmidrule(r){5-6}
		 & & EST & SE & EST & SE \\
		\midrule
Summer & $\lambda_1$ & 2.841 & 0.459 & 1.090 & 0.280 \\
   & $\lambda_0$ & 0.179 & 0.014 & 0.158 & 0.015 \\
   & $\sigma$ & 1.335 & 0.106 & 0.999 & 0.104 \\
   & $\sigma_\epsilon (\times 10^{-2})$ & 1.854 & 0.087 & 1.879 & 0.078 \\
  Winter & $\lambda_1$ & 6.225 & 0.825 & 4.720 & 0.711 \\
   & $\lambda_0$ & 0.118 & 0.010 & 0.114 & 0.009 \\
   & $\sigma$ & 1.506 & 0.095 & 1.454 & 0.089 \\
   & $\sigma_\epsilon (\times 10^{-2})$ & 0.908 & 0.036 & 0.934 & 0.043 \\
		\bottomrule
	\end{tabular}
	\endgroup
\end{table}

\section{Discussion}
\label{sec:disc}

Inactive periods and measurement errors are both necessary features of
animal movement models, if we are to successfully reconstruct
natural behaviors. Handling measurement errors in the MR
process is especially critical because, if discarded, a microscopic
amount of measurement error causes substantial bias in
estimation.
Our approach employing composite
likelihood is the first to make the MRME model practically feasible.
For movement data from predators that are known to have long inactive
periods \citep{jona:2007}, the MRME model has great potential
in revealing insights for
animal ecologists.  For example, the clear
seasonal patterns in movement patterns have immediate implications
for diverse ecological questions
\citep[e.g., seasonal foraging ecology, ][]{Elbroch:etal:2013}
and estimating animal abundance using models dependent upon
animal speed \citep{Moeller:etal:2018}.

The MRME model can be extended to meet further practical needs. For
some predators, an inactive period may have different purposes, such as
resting and food handling. \citet{Pozd:etal:2018}
introduced the moving-resting-handling (MRH) process that allows two
different types of motionless states.
The moving period may also represent different behaviors
\citep{Benh:2011, Kran:etal:2012}, and this can be
accommodated by allowing the volatility parameter to have
multiple levels. More generally, animal behavior
depends on a suite of intrinsic and extrinsic variables. Our MRME
model provides an important benchmark for building these
more realistic extensions.

\bibliographystyle{mcap}
\bibliography{mrme}

\appendix

\section{Formulas for High Dimensions}

Let $\X (t_k)$ and $\Z (t_k)$ be $d$-dimensional random vectors. Set
$\X (t_k) = (X_1(t_k), \dots, X_d(t_k))$,
$\mathbf{\epsilon_k} = (\epsilon_{1k}, \dots, \epsilon_{dk}) \sim \mathrm{MN}(\mathbf{0}, \sigma_{\epsilon}^2 \mathbf{I})$, and $\mathbf{\epsilon_i}$ and $\mathbf{\epsilon_j}$ are independent for $i \neq j$.
Then, $\Z(t_k) = (Z_1(t_k), \dots, Z_d(t_k)) = (X_1(t_k)+\epsilon_{1k}, \dots, X_d(t_k)+\epsilon_{dk})$.
The density $h_{ij}$ in the $d$-dimension case is given by
\[
  h_{10}(\x, t) = \int_0^t \prod_{i=1}^d \phi(x_i; \sigma^2 w)
  p_{10}(w, t) \dif w.
\]
Then we get that
\begin{align*}
	g_{10}(\z, t) &= \int_{\mathbb{R}} \dots \int_{\mathbb{R}} h_{10} (\z - \x ; \sigma^2 w)
	\prod_{i=1}^d \phi(x_i; 2\sigma_{\epsilon}^2) \dif x_1 \dots \dif x_d \\
	&= \int_{\mathbb{R}} \dots \int_{\mathbb{R}} \left[\int_0^t \prod_{i=1}^d
	\phi(z_i - x_i ; \sigma^2 w) p_{10}(w, t) \dif w\right]
	\prod_{i=1}^d \phi(x_i; 2\sigma_{\epsilon}^2) \dif x_1 \dots \dif x_d \\
	&= \int_0^t \prod_{i=1}^d \left[\int_{\mathbb{R}} \phi (z_i - x_i ; \sigma^2 w) \phi(x_i ;
	2\sigma_{\epsilon}^2) \dif x_i\right] p_{10}(w, t) \dif w.
\end{align*}
Similarly, we also have
\begin{align*}
	g_{00}(\z, t) =& \int_0^t \prod_{i=1}^d \left[\int_{\mathbb{R}} \phi (z_i - x_i ; \sigma^2 (t-w)) \phi(x_i ; 2 \sigma_{\epsilon}^2) \dif x_i\right] p_{00}(w, t) \dif w
	+ e^{-\lambda_0 t} \prod_{i=1}^d \phi (z_i, 2 \sigma_{\epsilon}^2), \\
	g_{01}(\z, t) =& \int_0^t \prod_{i=1}^d \left[\int_{\mathbb{R}} \phi (z_i - x_i ; \sigma^2 (t-w)) \phi(x_i ; 2\sigma_{\epsilon}^2) \dif x_i\right] p_{01}(w, t) \dif w, \\
	g_{11}(\z, t) =& \int_0^t \prod_{i=1}^d \left[\int_{\mathbb{R}} \phi (z_i - x_i ; \sigma^2 w) \phi(x_i ; 2\sigma_{\epsilon}^2) \dif x_i\right] p_{11}(w, t) \dif w\\
                    & + e^{-\lambda_1 t} \prod_{i=1}^d \left[\int_{\mathbb{R}} \phi (z_i - x_i ; \sigma^2 t) \phi(x_i ; 2\sigma_{\epsilon}^2) \dif x_i\right].
\end{align*}
Let us  mention that these formulas do not require numerical multiple integral
evaluation and, as a consequence, are computationally efficient.

\label{lastpage}

\end{document}